\documentclass[]{article}

\usepackage{amsfonts,amsbsy,natbib}

\renewcommand{\vec}[1]{\boldsymbol{#1}}

\begin{document}
  
{\noindent\bf THERMODYNAMICS AND PREFERRED FRAME}
\vspace{\baselineskip}

\hspace*{4pc}
\begin{tabular}{l}
{\bf Jakub Rembieli{\'n}ski,
Kordian Andrzej Smoli{\'n}ski,}\\
{\bf Grzegorz Duniec}\\
\\
{\it Department of Theoretical Physics, University of Lodz,}\\
{\it ul.~Pomorska 149/153, 90-236 {\L}{\'o}d{\'z}, Poland}\\
{\it jaremb@krysia.uni.lodz.pl, smolin@krysia.uni.lodz.pl}
\end{tabular}
\vspace{3\baselineskip}





{\noindent The Lorentz covariant statistical physics and
  thermodynamics is formulated within the preferred frame approach.
  The transformation laws for geometrical and mechanical quantities
  such as volume and pressure as well as the Lorentz-invariant measure
  on the phase space are found using Lorentz transformations in
  absolute synchronization.  Next, the probability density and
  partition function are investigated using the preferred frame
  approach, and the transformation laws for internal energy, entropy,
  temperature and other thermodynamical potentials are established.
  The Lorentz covariance of basic thermodynamical relations, including
  Clapeyron's equation and Maxwell's relations is shown.  Finally, the
  relation of presented approach to the previous approaches to
  relativistic thermodynamics is briefly discussed.}
\vspace{\baselineskip}



\vspace{2\baselineskip}
{\noindent\bf 1. INTRODUCTION}
\vglue\baselineskip

\noindent 
The question of relativistic formulation of thermodynamics is almost
as old as Special Relativity.  The first attempt for the formulation
of relativistic thermodynamics was done by Einstein, Planck and von
Laue \cite{einstein07,laue61} in 1907{\textendash}1908.  They concentrated on
establishing of transformation laws of thermodynamical quantities by
defining the transformation properties of work and heat.  The main
results of these works were the transformation laws for the change of
entropy $dS$, temperature $T$ and heat transfer $\delta Q$
\begin{equation}
  \label{eq:planck}
  dS = dS_0, \quad T = T_0 \sqrt{1 - v^2}, \quad \delta Q = \delta Q_0 \sqrt{1 -
   v^2},
\end{equation}
where the index $0$ denotes values taken in the center of mass system
of the gas, and $v$ is the velocity of the observer with respect to
the center of mass system of the gas (hereafter we work in the natural
system of units, where $c = 1$).

The derivation of transformation laws of thermodynamical quantities by
Einstein and Planck was questioned in 60's by Ott \cite{ott63} and
Arzeli{\`e}s \cite{arzelies65}.  Their argumentation led to the
transformation laws opposite to (\ref{eq:planck}), namely
\begin{equation}
  \label{eq:ott}
  \delta Q = \frac{\delta Q_0}{\sqrt{1 - v^2}}, \quad T = \frac{T_0}{\sqrt{1 - v^2}}
\end{equation}
(with the same notation as above).  We refer the Reader to the review
by ter~Haar and Wergeland \cite{haar71} for extensive discussion
regarding the formulation of relativistic thermodynamics by Planck and
Ott.

Moreover, Landsberg \cite{landsberg66,landsberg68} had questioned the
traditional relativistic generalization of the thermodynamics law and
came into conclusion that temperature and heat transfer are invariant
\begin{equation}
  \label{eq:landsberg}
  T = T_0, \quad \delta Q = \delta Q_0
\end{equation}
(internal energy $dU = dU_0$ and pressure $p = p_0$ are invariant in
Landsberg's approach, too).

In this paper we look at this old-standing problem from a completely
new point of view.  Our approach is motivated by the fact that the
proper formulation of quantum mechanics needs a preferred frame
\cite{bell81}.  In this way one avoids many serious interpretational
problems in quantum mechanics
\cite{caban99,hardy92,percival98,percival00,rembielinski97}.
Moreover, Einstein himself had accepted the existence of such a
``non-mechanical'' preferred frame \cite{einstein22}.  We would like
to point out that the notion of preferred frame has nothing to do with
the obsolete notion of ``ether''.

We will study consequences of the absolute synchronization scheme
\cite{caban99,rembielinski80,rembielinski97} and the existence of a
preferred frame for thermodynamics and statistical physics.  As a
result we will find the transformation laws for thermodynamical
quantities with as few as possible changes in the thermodynamical
relations.  It is remarkable that in the simplest physical situations
the transformation law for the temperature is exactly the formula
found by Einstein and Planck (\ref{eq:planck}).

The paper is organized as follows.  In the next section we sketch
basic properties of the kinematics in absolute synchronization{\textemdash}namely
we state the Lorentz transformations and show how geometrical and
mechanical quantities transform in absolute synchronization.  In
Section~3 we construct the probability density and partition function
in absolute synchronization approach, then we derive transformation
properties for basic thermodynamical quantities, such as internal
energy, entropy and temperature.  Section~4 discusses some basic
thermodynamical relations and shows their Lorentz covariance in
absolute synchronization approach.  The last section contains final
remarks and conclusions.

\vspace{2\baselineskip}
{\noindent\bf 2. KINEMATICS IN ABSOLUTE SYNCHRONIZATION}
\vglue\baselineskip

{\noindent\bf 2.1. Lorentz transformations in absolute synchronization}
\vglue\baselineskip

\noindent
In this section we briefly describe main results related to the
absolute synchronization scheme.  Derivation of these results can be
found in \cite{rembielinski80,rembielinski97}.  The main idea is based
on the well known fact that the definition of the time coordinate
depends on the procedure used to synchronize clocks.  The choice of
this procedure is a convention
\cite{anderson98,croca99,jammer79,mansouri77,reichenbach69,will92,will93}
(this fact is known as ``conventionality thesis'').  Therefore, the
form of Lorentz transformation depends on the synchronization scheme,
and we can find a synchronization procedure which leads to the desired
form of Lorentz transformation.  To perform such a choice under
assumed absolute time synchronization we have to distinguish, at least
formally, one inertial frame.  Such a frame is called preferred frame
\cite{caban99,rembielinski80,rembielinski97}; it can be possibly
identified with the cosmic background radiation frame.  The
four-velocity of the preferred frame with respect to the observer in
an inertial frame ${\cal O}_u$ is denoted by $u^\mu$.

In absolute synchronization the transformation law of {\em
  contravariant\/} components of coordinates between inertial frames
is expressed by the formula \cite{caban98,caban99,rembielinski97}
\begin{equation}
  \label{eq:CT-cont}
  [x'(u')]^\mu = [D(\Lambda, u)]^\mu{}_\nu [x(u)]^\nu,
\end{equation}
where $\Lambda$ is an element of Lorentz group, $u$ is a four-velocity and
$D(\Lambda, u)$ is $4\times4$ matrix depending on $\Lambda$ and $u$.  Transformation
law for the contravariant components of four-velocity $u^\mu$ is
\begin{equation}
  \label{eq:u-cont}
  u'{}^\mu = [D(\Lambda, u)]^\mu{}_\nu u^\nu.
\end{equation}

Matrix $D(\Lambda, u)$ for any rotation $R\in\mathrm{SO}(3)$ has the form
\begin{equation}
  \label{eq:D-mat:R}
  D(R, u) = \left(
    \begin{array}{c|c}
      1 & 0\\
      \hline
      0 & R
    \end{array}
  \right),
\end{equation}
while for the boosts $w$ is given by
\begin{equation}
  \label{eq:D-mat}
  D(w, u) = \left(
    \begin{array}{c|c}
      \frac1{w^0} & 0\\
      \hline
      -\vec{w} & I + \frac{\vec{w}\otimes\vec{w}^T}{1 + \sqrt{1 + (\vec{w})^2}} 
      - u^0 \vec{w}\otimes\vec{u}^T
    \end{array}
  \right),
\end{equation}
where $w^\mu$ is the four-velocity of the system ${\cal O}_{u'}$ with
respect to the system ${\cal O}_u$ (here $\otimes$ denotes the direct
product of matrices).  We can express the four-velocity $w^\mu$ by the
four-velocities of the preferred frame $u$ and $u'$ observed from the
system ${\cal O}_u$ and ${\cal O}_{u'}$ respectively
\begin{equation}
  \label{eq:w-uu}
  w^0 = \frac{u^0}{u'{}^0}, \quad \vec{w} = \frac{(u^0 + u'{}^0) (\vec{u}
    - \vec{u}')}{1 + u^0 u'{}^0 (1 + \vec{u}\cdot\vec{u}')}.
\end{equation}
The components $w^\mu$ can be expressed in terms of the velocity
$\vec{v} = \vec{w}/w^0$ of the frame ${\cal O}_{u'}$ with respect
to ${\cal O}_u$ via the relation
\begin{equation}
  \label{eq:vw}
  w^0 = \frac1{\sqrt{(1 + u^0 \vec{u}\cdot\vec{v})^2 - (\vec{v})^2}}.
\end{equation}

Now the transformation law for the {\em covariant\/} components of
coordinates reads
\begin{equation}
  \label{eq:Dt-mat}
  (D^{-1}(w, u))^T = \left(
    \begin{array}{c|c}
      w^0 & \vec{w}^T\\
      \hline
      0 & I - \frac{\vec{w}\otimes\vec{w}^T}{1 + \sqrt{1 + (\vec{w})^2}} + 
      \frac{u^0}{w^0} \vec{u}\otimes\vec{w}^T
    \end{array}
  \right).
\end{equation}

Comparing the transformation laws for the contravariant and covariant
components of coordinates we see that under the transformation
contravariant components of spatial coordinates mix with the time
coordinate, while covariant do not.  Notice also that the time
coordinate $x^0$ is rescaled by a positive factor only.  Indeed, for
Lorentz boosts we can rewrite (\ref{eq:CT-cont}) using
(\ref{eq:D-mat}) as follows
\begin{equation}
  \label{eq:x-cont}
  x'{}^0 = \frac1{w^0} x^0, \quad \vec{x}' = -\vec{w} x^0 + M \vec{x},
\end{equation}
where
\begin{equation}
  \label{eq:Mt}
  M = I + \frac{\vec{w}\otimes\vec{w}^T}{1 + \sqrt{1 + (\vec{w})^2}} - u^0
  \vec{w}\otimes\vec{u}^T;
\end{equation}
while (\ref{eq:Dt-mat}) gives
\begin{equation}
  \label{eq:x-cov}
  x'_0 = w^0 x_0 + \vec{w}\cdot\underline{\vec{x}}, \quad
  \underline{\vec{x}}' = \underline{M}\, \underline{\vec{x}},
\end{equation}
where
\begin{equation}
  \label{eq:M}
  \underline{M} = I - \frac{\vec{w}\otimes\vec{w}^T}{1 + \sqrt{1 +
      (\vec{w})^2}} + \frac{u^0}{w^0} \vec{u}\otimes\vec{w}^T
\end{equation}
and $\underline{\vec{x}}$ denotes the {\em covariant position
  three-vector}.  The above transformation laws apply to contravariant
and covariant components of any tensor in the absolute
synchronization.  It is also important to stress that the time
transformation law enables us to introduce the notion of absolute
causality within this framework.

The line element
\begin{equation}
  \label{eq:ds}
  ds^2 = g(u)_{\mu\nu} dx^\mu dx^\nu
\end{equation}
is invariant under Lorentz transformations if the metric tensor is of
the following form
\begin{equation}
  \label{eq:g}
  [g(u)_{\mu\nu}] = \left(
    \begin{array}{c|c}
      1 & u^0 \vec{u}^T\\
      \hline
      u^0 \vec{u} & -I + (u^0)^2 \vec{u}\otimes\vec{u}^T
    \end{array}
  \right);
\end{equation}
while its contravariant counterpart is
\begin{equation}
  \label{eq:g-1}
  [g(u)^{\mu\nu}] = \left(
    \begin{array}{c|c}
      (u^0)^2 & u^0 \vec{u}^T\\
      \hline
      u^0 \vec{u} & -I
    \end{array}
  \right).
\end{equation}
From (\ref{eq:g-1}) it follows that the geometry of the three-space is
Euclidean, i.e.\ $dl^2 = d\vec{x}^2$.

Note also that in every inertial frame $u_\mu = (u_0,
\underline{\vec{0}})$ (this follows immediately from $u_\mu = g_{\mu\nu}
u^\nu$, the relation $u^\mu u_\nu = 1$ and (\ref{eq:g-1})), so the condition
$u^\mu u_\mu = 1$ turns into the relation
\begin{equation}
  \label{eq:u0u0}
  u^0 u_0 = 1.
\end{equation}

Finally the relationship of $x^\mu$ with coordinates in the standard
(Einstein--Poincar{\'e}) synchronization scheme is given by
\begin{equation}
  \label{eq:EP}
  x^0_E = x^0 - u^0 \vec{u}\cdot\vec{x}, \quad \vec{x}_E = \vec{x},
\end{equation}
so the time lapses in the same point of the space are identical in
both synchronizations.  From (\ref{eq:EP}) we are able to derive also
the relationship between the velocities in absolute and standard
synchronizations \cite{rembielinski97}.

A more exhaustive discussion of the absolute synchronization, and its
geometrical interpretation in terms of frame bundles is given in
\cite{caban99,rembielinski97}.

\vspace{\baselineskip}
{\noindent\bf 2.2. Transformation properties of volume}
\vglue\baselineskip

\noindent 
The (three-) volume of the infinitesimal cube is defined by $dV = dx^1
\land dx^2 \land dx^3 \equiv d^3\vec{x}$ where all the $dx$'s are taken in the same
time, i.e.\ with $dx^0 = 0$.  Note that the condition $dx^0 = 0$ is
invariant under the transformations (\ref{eq:CT-cont}), because if
$dx^0 = 0$ in the frame of the observer ${\cal O}_u$, it follows from
(\ref{eq:x-cont}) that $dx'{}^0 = 0$ for any other observer ${\cal
  O}_{u'}$.

To find the transformation law for $dV$ first consider the invariant
quantity $u\, d\sigma \equiv u_\mu d\sigma^\mu$, where $d\sigma^\mu = \frac{1}{3!}  \epsilon^{\mu\nu\kappa\lambda}
dx_\nu \land dx_\kappa \land dx_\lambda$ is an element of hyper-surface,
\begin{equation}
  \label{eq:dsig0}
  u\, d\sigma = u_0\, d\sigma^0 = \frac{\epsilon^{0ijk}}{3!} u_0\, dx_i \land dx_j \land dx_k \equiv
  u_0\, d^3\underline{\vec{x}}.
\end{equation}
Using the metric tensor $g^{\mu\nu}(u)$ we may rewrite $dx_i$'s in terms
of $dx^\mu$'s and for $dx^0 = 0$ we obtain
\begin{equation}
  \label{eq:dsig0-det}
  u_0\, d\sigma^0 = u_0 \det\left(-I + (u^0)^2 \vec{u}\otimes\vec{u}^T\right)
  d^3\vec{x} = -u^0\, dV.
\end{equation}
Therefore $dV = -\frac{1}{u^0} u\, d\sigma$ transforms according to the
Eq.~(\ref{eq:x-cont}) as follows
\begin{equation}
  \label{eq:dV'}
  dV' = w^0\, dV,
\end{equation}
so
\begin{equation}
  \label{eq:V'}
  V' = w^0 V,
\end{equation}
where $V = \int_{t = {\rm const}}dV$.

\vspace{\baselineskip}
{\noindent\bf 2.3. Invariant measure on the phase space}
\vglue\baselineskip

\noindent
Now we construct an invariant measure on the phase space of a free
particle.

Let us begin with an invariant measure in momentum space, assuming
proper spectral conditions for the particle four-momentum, namely $p^0
> 0$ and $p^2 = m^2$ (see \cite{caban99}):
\begin{equation}
  \label{eq:m-mom}
  d\mu(p) \equiv \theta(p^0) \delta(p^2 - m^2)\, d^4p,
\end{equation}
i.e.\ 
\begin{equation}
  \label{eq:dmu}
  d\mu(p) = \frac{d^3 \underline{\vec{p}}}{2 p^0},
\end{equation}
where $\underline{\vec{p}}$ is {\em covariant momentum three-vector}.
Multiplying (\ref{eq:dmu}) by the invariant $2 (u_\mu p^\mu)$ we obtain an
invariant measure on the phase space with proper scaling behavior:
\begin{equation}
  \label{eq:m-mom-dp}
  2 u p\, d\mu(p) = u_0\, d^3\underline{\vec{p}}.
\end{equation}

Finally, taking into account (\ref{eq:dsig0-det}), we obtain the
following invariant measure on a single particle phase space
\begin{equation}
  \label{eq:dG}
  d\Gamma = 2 d\mu(p)\, u_\mu d\sigma^\mu\, u_\nu p^\nu\,  
  = -d^3 \vec{x}\, d^3\underline{\vec{p}}.
\end{equation}
We see that $d\Gamma$ has the form analogous to the non-relativistic case.

\vspace{\baselineskip}
{\noindent\bf 2.4. Pressure}
\vglue\baselineskip

\noindent
The pressure $P$ of perfect fluid can be defined by the relation to
its stress-energy tensor, namely by
\begin{equation}
  \label{eq:stress}
  \sigma^{\mu\nu} = (\rho + P) \mathfrak{w}^\mu \mathfrak{w}^\nu 
  - P g^{\mu\nu},
\end{equation}
where $\rho$ is the fluid density and $\mathfrak{w}^\mu$ is the
four-velocity of fluid with respect to the observer ${\cal O}_u$.
Pressure is then invariant
\begin{equation}
  \label{eq:P'}
  P' = P
\end{equation}
and can be separated from the density using invariant relations
\begin{equation}
  \label{eq:traces}
  \mathrm{Tr}\sigma \equiv \sigma^{\mu\nu} g_{\mu\nu} = \rho + 3 P, 
  \quad \mathrm{Tr}\sigma^2 = \rho^2 + 3 P^2.
\end{equation}

Summarizing this section, we conclude that in the absolute
synchronization scheme the volume and pressure have simple
transformation laws (\ref{eq:V'}) and (\ref{eq:P'}), while the
invariant measure on phase space has the very suggestive form
(\ref{eq:dG}).

\vspace{2\baselineskip}
{\noindent\bf 3. STATISTICAL MECHANICS AND ABSOLUTE SYNCHRONIZATION}
\vglue\baselineskip

{\noindent\bf 3.1. Probability density and partition function}
\vglue\baselineskip

\noindent 
Consider the perfect gas of $N$ non-interacting particles.  We start
with the following formula for a probability density of finding the
$i$-th particle in an infinitesimal element of phase space $d\Gamma_i$
\begin{equation}
  \label{eq:prob1-n}
  \frac1{Z_i} \exp\left(-\beta (u^0)^n p^0_{(i)}\right),
\end{equation}
which is the natural modification of the Boltzmann--Maxwell
distribution.  Here $\beta = \frac1{k T}$, $n$ is an arbitrary power to be
fixed later, $Z_i$ is the normalization factor---partition function
\begin{equation}
  \label{eq:intZ1-n}
  Z_i(\beta, V) = \int d\Gamma_i\, \exp\left(-\beta (u^0)^n p^0_{(i)}\right).
\end{equation}
The factor $(u^0)^n$ is introduced to preserve the invariance of the
argument of the exponential function under transformations
(\ref{eq:CT-cont}).

Therefore the probability distribution of finding $N$ non-interacting
identical particles of the gas in an element of the phase space $d\Gamma_1
\cdots d\Gamma_N$ is
\begin{equation}
  \label{eq:probN-n}
  \frac1Z \exp\left(-\beta (u^0)^n \sum_{i=1}^N p^0_{(i)}\right),
\end{equation}
where the partition function $Z(\beta, V)$ for the gas is
\begin{eqnarray}
  Z(\beta, V) &=& \int \cdots \int d\Gamma_1 \cdots d\Gamma_N\, 
  \exp\left(-\beta (u^0)^n \sum_{i=1}^N p^0_{(i)}\right)\nonumber\\
  &=& \left(V \int d^3\underline{\vec{p}}
    \exp\left(-\beta (u^0)^n p^0\right)\right)^N.
  \label{eq:intZ-n}
\end{eqnarray}

Of course the partition function must be an invariant of the Lorentz
transformations (\ref{eq:M})
\begin{equation}
  \label{eq:Z'}
  Z' = Z.
\end{equation}
This requires that the argument of the exponential in
(\ref{eq:intZ-n}) also must be an invariant.  Since $p'{}^0 =
\frac1{w^0} p^0$ and $u'{}^0 = \frac1{w^0} u^0$ it follows that $\beta' =
(w^0)^{n+1} \beta$, i.e.\ 
\begin{equation}
  \label{eq:beta}
  T' = \frac1{(w^0)^{n+1}} T.
\end{equation}

\vspace{\baselineskip}
{\noindent\bf 3.2. Internal energy}
\vglue\baselineskip

\noindent
In Lorentz covariant dynamics with absolute synchronization
Hamiltonian, as the generator of the time translations, is identified
with 0-th covariant component of four-momentum: $H_i \equiv {p_0}_{(i)}$.
Therefore energy of the system of $N$ non-interacting particles is
equal to
\begin{equation}
  \label{eq:H}
  H = \sum_{i=1}^N {p_0}_{(i)}.
\end{equation}
The internal energy is defined as the mean value of Hamiltonian, i.e.,
\begin{eqnarray}
  U = \langle H\rangle 
  &=& \frac1Z \int \cdots \int d\Gamma_1 \cdots d\Gamma_N\, \left(\sum_{k=1}^N {p_0}_{(k)}\right) 
  \exp\left(-\beta (u^0)^n \sum_{i=1}^N p^0_{(i)}\right)\nonumber\\
  \label{eq:U}
  &=& \frac1Z \left(V \int d^3\underline{\vec{p}}\, 
    p_0 \exp\left(-\beta (u^0)^n p^0\right)\right)^N,
\end{eqnarray}
where $p_0$ and $p^0$ are covariant and contravariant components of
the momentum of a single particle of the gas under consideration,
respectively.

On the other hand in statistical physics internal energy is related to
the partition function by
\begin{equation}
  \label{eq:dZ/db}
  U = -\frac{\partial\ln Z}{\partial\beta} \equiv -\frac1Z \frac{\partial Z}{\partial\beta}.
\end{equation}
Taking into account Eq.~(\ref{eq:intZ-n}) we obtain from
(\ref{eq:dZ/db}) that
\begin{equation}
  \label{eq:U-Z}
  U = \frac1Z \left((u^0)^{n+2} V \int d^3\underline{\vec{p}}\, p_0 
  \exp\left(-\beta (u^0)^n p^0\right)\right)^N.
\end{equation}

Comparison of (\ref{eq:U}) and (\ref{eq:U-Z}) fixes the power of
$u^0$, so we have to choose $n = -2$.  Therefore $\beta' = \beta/w^0$
and in consequence
\begin{equation}
  \label{eq:U'}
  U' = w^0 U
\end{equation}
and
\begin{equation}
  \label{eq:T'}
  T' = w^0 T.
\end{equation}

Therefore the probability distribution (\ref{eq:probN-n}) finally takes the
form
\begin{equation}
  \label{eq:probN}
  \frac1Z \exp\left(-u_0 \beta \sum_{i=1}^N u p_{(i)}\right),
\end{equation}
while the partition function is
\begin{equation}
  \label{eq:intZ}
  Z(\beta, V) = \left(V \int d^3\underline{\vec{p}}\, e^{-u_0 \beta u p}\right)^N,
\end{equation}
where we have used (\ref{eq:u0u0}) and $u p \equiv u_\mu p^\mu = u_0 p^0$.

Notice, that in the preferred frame ($u^0 =1$, $\vec{u} = 0$) $Z$
takes the standard form.

\vspace{\baselineskip}
{\noindent\bf 3.3. Entropy}
\vglue\baselineskip

\noindent
In fixed external conditions every simple thermodynamical system is
uniquely described by its internal energy $U$ and volume $V$.  The
entropy $S = S(U, V)$ is a unique function of these parameters and is
maximal for equilibrium states of the system.

We treat entropy as the measure of the information on the system.
According to this approach we use Boltzmann definition of entropy
\begin{equation}
  \label{eq:S-def}
  S(U, V) = -\sum_{i=1}^N \rho_i \ln \rho_i,
\end{equation}
where $\rho_i$ are probabilities of finding the $i$-th particle in a
given microscopic event realizing the macroscopic state of the system
described by energy $U$ and volume $V$.

If the observer staying in the frame ${\cal O}_u$ describes the
system under consideration by $U$, $V$, the other observer, staying in
the frame ${\cal O}_{u'}$ describes the system by $U'$, $V'$.
Entropy $S'$ in the frame ${\cal O}_{u'}$ is equal to
\begin{equation}
  \label{eq:S'-def}
  S'(U', V') = -\sum_{i=1}^N \rho_i \ln\rho_i,
\end{equation}
because the macroscopic state $(U', V')$ is realized by the same
microscopic events.  So, entropy is invariant
\begin{equation}
  \label{eq:S'}
  S' = S.
\end{equation}

Note that using (\ref{eq:Z'}), (\ref{eq:U'}), and (\ref{eq:T'}) we
obtain the same transformation law (\ref{eq:S'}) on a basis of the
relation between entropy and partition function
\begin{equation}
  \label{eq:S-Z}
  S = k \ln Z + \frac{U}T.
\end{equation}

\vspace{2\baselineskip}
{\noindent\bf 4. LAWS OF THERMODYNAMICS AND ABSOLUTE SYNCHRONIZATION}
\vglue\baselineskip

\noindent
We can easily check that the usual laws of statistical mechanics and
thermodynamics are consistent with the transformation law of physical
quantities derived in sections~2 and~3.

The transformation law for pressure (\ref{eq:P'}) is consistent with
the statistical definition of pressure
\begin{equation}
  \label{eq:P_Z}
  P = k T \left(\frac{\partial\ln Z}{\partial V}\right)_T,
\end{equation}
provided the transformation laws for volume (\ref{eq:dV'}), partition
function (\ref{eq:Z'}) and temperature (\ref{eq:T'}) hold.

The form of the partition function of the perfect gas (\ref{eq:intZ})
and the above equation imply the relation
\begin{equation}
  \label{eq:Clapeyron}
  P V = N k T,
\end{equation}
i.e.\ the usual Clapeyron's equation of the state of the perfect gas.

Moreover, the relation $P' = P$ agrees with the thermodynamical
definition of pressure (Maxwell's equation)
\begin{equation}
  \label{eq:P_th}
  P = -\left(\frac{\partial U}{\partial V}\right)_S.
\end{equation}

Maxwell's equation for temperature is covariant in the absolute
synchronization, too.  Indeed, consider the relation
\begin{equation}
  \label{eq:T-def}
  T = \left(\frac{\partial U}{\partial S}\right)_V
\end{equation}
and use (\ref{eq:U'}) and (\ref{eq:S'}) to obtain $T' = w^0 T$ again.

The above relation allows us to keep the first law of thermodynamics
unchanged:
\begin{equation}
  \label{eq:1st_law}
  dU = T dS - P dV.
\end{equation}

Therefore we can define the heat
\begin{equation}
  \label{eq:dQ}
  \delta Q = T dS
\end{equation}
and work
\begin{equation}
  \label{eq:dL}
  \delta L = -P dV.
\end{equation}
It is easy to check that $\delta Q' = w^0 \delta Q$ and $\delta L' = w^0 \delta L$.

Transformation laws for the thermodynamical parameters are collected
in Table~\ref{tab:potentials}.
\begin{table*}[t]
  \caption{Transformation laws for thermodynamical parameters using 
    absolute synchronization; $w^0 = 1/\sqrt{(1 + u^0 \vec{u}\cdot\vec{v})^2 - 
      (\vec{v})^2}$.\vspace{\baselineskip}}
  \label{tab:potentials}
  \begin{tabular}{llc}
    \hline\hline
    Parameter & Notation and definition & 
    \multicolumn{1}{l}{Transformation law}\\ 
    \hline
    Volume & $V$ & $V' = w^0 V$\\
    Temperature & $T$ & $T' = w^0 T$\\
    Internal energy & $U(S, V)$ & $U' = w^0 U$\\
    Enthalpy & $H(S, P) = U + P V$ & $H' = w^0 H$\\
    Helmholtz free energy & $F(T, V) = U - T S$ & $F' = w^0 F$\\
    Gibbs' free energy & $G(T, P) = U - T S + P V$ & $G' = w^0 G$\\
    \hline
    Pressure & $P$ & $P' = P$\\
    Entropy & $S(U, V)$ & $S' = S$\\
    Partition function & $Z(T, V)$ & $Z' = Z$\\
    Massieu's potential & $\Psi(1/T, V) = -F/T$ & $\Psi' = \Psi$\\
    Planck's potential & $\Phi(1/T, V) = -G/T$ & $\Phi' = \Phi$\\
    \hline\hline
  \end{tabular}
  \vspace{\baselineskip}
\end{table*}
The fundamental thermodynamics relations that are covariant under
Lorentz group in absolute synchronization scheme are shown in
Table~\ref{tab:relations}.
\begin{table}
  \caption{Thermodynamical relations covariant under Lorentz group in 
    absolute synchronization.\vspace{\baselineskip}}
  \label{tab:relations}
  \begin{tabular}{ccc}
    \hline\hline
    $U = k T^2 (\partial\ln Z/\partial T)$, & $S = k \ln Z + U/T$, &
    $P = k T (\partial\ln Z/\partial V)_T$, \\ 
    $P V = N k T$, & $P = -(\partial U/\partial V)_S$, & $T = (\partial U/\partial S)_V$,\\
    $dU = T\, dS - P\, dV$, & $\delta Q = T\, dS$, & $\delta L = -P\, dV$\\
    \hline\hline
  \end{tabular}
  \vspace{\baselineskip}
\end{table}

\vspace{2\baselineskip}
{\noindent\bf 5. CONCLUSIONS}
\vglue\baselineskip

\noindent
In this paper we have formulated Lorentz-covariant thermodynamics.
Such a formulation can be possible due to use of a preferred frame and
absolute synchronization.  As it was suggested in
\cite{caban99,rembielinski80,rembielinski97}, the preferred frame can
be possibly identified with the cosmic background radiation frame.

We have defined the distribution density and partition function that
transforms covariantly under Lorentz boost in the framework of
absolute synchronization.  This is possible because the absolute
synchronization allows for the existence of invariant measure on the
phase space of a one-particle system.  On this basis we derive
transformation properties of all thermodynamical quantities, including
entropy, temperature, internal energy and pressure.  We can also check
that some basic thermodynamical relations such as thermodynamical
relation for pressure, Maxwell's relations, and Clapeyron's equation
are Lorentz covariant in the absolute synchronization scheme.

We would like to point out that, in contrary to the transformation
laws found by Einstein, Planck and von Laue \cite{einstein07,laue61}
or the ones found by Ott and Arzeli{\`e}s \cite{arzelies65,ott63}, the
preferred frame approach allows one to find the transformation laws
for thermodynamical quantities in any two inertial frames, not only
between a given inertial frame and the system of center of mass of the
gas.  Moreover, it is easy to verify that in the simplest physical
situations: (i)~when the observer is in the preferred frame, and
(ii)~when the gas is in the preferred frame, the transformation law
(\ref{eq:T'}) becomes\footnote{To compare transformation laws in
  absolute and standard synchronizations we must replace the
  velocities in absolute synchronization by the velocities in the
  standard (Einstein{\textendash}Poincar{\'e}) synchronization (see
  \cite{rembielinski97}).}  the Einstein{\textendash}Planck law~(\ref{eq:planck}).

\vspace{2\baselineskip}
{\noindent\bf ACKNOWLEDGEMENTS}
\vglue0pt
\vspace{\baselineskip}
\noindent  
This work is supported by the University of Lodz grant.


\end{document}